\documentclass[prd,aps,floats,twocolumn,nofootinbib,superscriptaddress]{revtex4-1}
\usepackage{slashed}
\usepackage{mathtools}
\usepackage{amsfonts}
\usepackage{amssymb}
\usepackage{epsfig}
\usepackage{rotating}
\usepackage{url}
\usepackage{times}
\usepackage{color}
\usepackage{bm}
\usepackage{xcolor,colortbl}
\usepackage{hyperref}
\usepackage[alsoload=hep]{siunitx}
\usepackage{enumitem}
\usepackage{fancyhdr}
\usepackage{anyfontsize}
\usepackage{wasysym}
\usepackage{empheq}

\newcommand{\nn}{\nonumber}

\newcommand{\be}{\begin{equation}}
\newcommand{\ee}{\end{equation}}
\newcommand{\bea}{\begin{eqnarray}}
\newcommand{\eea}{\end{eqnarray}}

\begin{document}


\title{Linking the Baum-Hawking-Coleman Mechanism with Unimodular Gravity and Vilenkin's Probability Flux}

\author{James Page}
\email{j.page@sussex.ac.uk}
\affiliation{Experimental Particle Physics Group, Department of Physics and Astronomy, Pevensey II Building, University of Sussex, Falmer, Brighton, BN1 9QH, United Kingdom}
\affiliation{Theoretical Physics Group, The Blackett Laboratory, Imperial College, Prince Consort Rd., London, SW7 2BZ, United Kingdom}
\author{Jo\~{a}o Magueijo}
\email{j.magueijo@imperial.ac.uk}
\affiliation{Theoretical Physics Group, The Blackett Laboratory, Imperial College, Prince Consort Rd., London, SW7 2BZ, United Kingdom}
\date{\today}

\begin{abstract}
We revisit a mechanism proposed by Hawking to resolve the cosmological constant problem
(and the controversy it generated) to identify possibly more palatable alternatives and explore new connections and interpretations. 
In particular, through the introduction of a new action coupling the four-form field strength $F = dA$ to the cosmological constant via a dynamical field $\lambda (x)$, a novel Baum-Hawking-Coleman type mechanism is presented. 
This mechanism can be seen as a generalisation of Unimodular Gravity. A theory with a similar
coupling to ``$F^2$'' is also presented, with promising results. We show how in such theories the 3-form is closely related to the Chern-Simons density, and its associated definition of time. 
On the interpretational front, we propose a method avoiding the standard Euclidean action prescription,
which makes use of Vilenkin's probability flux. 
\end{abstract}

\maketitle


\section{Introduction and background to the problem}

\subsection{Hawking's Original Proposal}

The fine tuning of the cosmological constant $\Lambda$ is infamously one of the greatest mysteries of modern Physics~\cite{weinberg1989, kaloper2014sequestering, bousso2000quantization, bousso2008cosmological, Padilla, enz1960nullpunktsenergie}. Attempts to resolve this problem have motivated many attempts at modifications of General Relativity (GR). One example is Unimodular Gravity~\cite{Unruh:1988in, henneaux1989, eichhorn2013unimodular}, whereby the metric determinant is constrained to $|g| = 1$. The point is to force the vacuum contributions of the stress-energy tensor to drop out of the equations of motion (EoM). However, this only replaces such contributions by 
an integration constant, taking on the role of $\Lambda$, leaving the problem unresolved~\cite{Padilla,Padilla:2014yea}.

In a more promising approach, Hawking proposed~\cite{hawking1984} the introduction of a gauge three-form $A$,
so that the Euclidean gravity action is:
\begin{equation}
    S = \frac{\kappa}{2} \int \epsilon_{abcd} e^a e^b \left( R^{cd} + \frac{1}{6} e^c e^d \Lambda \right) + \int F \wedge \ast F,
    \label{1}
\end{equation}
where $F$ is the field strength of $A$ ($F=dA$), $e^a = e^a_{\mu} dx^{\mu}$ with $\{e^a_{\mu}\}$ vierbein \cite{deser1976canonical, kibble1961lorentz, nakahara2003geometry}, $R^{ab} = R^{ab}_{\;\; \mu \nu} dx^{\mu} dx^{\nu}$ the curvature two-form, and $\kappa = (16 \pi G)^{-1}$ with $G$ Newton's gravitation constant. The EoM for just $A$ is then solved, with the familiar result
\begin{equation}
    d \ast F = 0 \;\; \Rightarrow \;\; \ast F = c,
    \label{eom1}
\end{equation}
a constant. When substituted back into the action, this yields 
\begin{equation}
    S_{\text{eff}} = - \frac{48 \pi^2 \kappa}{\lambda_{eff}(c)} \; , \; \; \lambda_{eff} (c) = \Lambda + \frac{c^2}{2 \kappa},
    \label{1b}
\end{equation}
for a closed universe (with Hartle-Hawking boundary conditions), where the spacetime volume $V = \int dx^4 \sqrt{|g|} = \int \ast 1 / 4!$ is $V = 24 \pi^2 / \lambda_{eff}(c)^2$ \cite{weinberg1989}. One can then use this to make a probability argument via the path integral formulation of quantum field theory (QFT). In this framework, the probability amplitude of going from state 1 (with a field configuration $\phi_1$ at time $t_1$), to a state 2 (likewise $\phi_2$ at $t_2$) 
\begin{equation}
    \langle\phi_2(t_2)|\phi_1(t_1)\rangle = \int \mathcal{D}[\phi] e^{i S_L[\phi]},
    \label{QFT1}
\end{equation}
where $S_L$ is the Lorentzian action, and the measure $\mathcal{D}[\phi]$ indicates the integral over all field configurations with $\phi$ equal to $\phi_1$ and $\phi_2$ at times $t_1$ and $t_2$ respectively. Since the integral oscillates rather than converging, a Wick rotation is often performed, taking the integral over imaginary time $\tau = it$, becoming
\begin{equation}
    \langle\phi_2(t_2)|\phi_1(t_1)\rangle = \int \mathcal{D}[\phi] e^{-S_E[\phi]},
    \label{QFT2}
\end{equation}
a converging path integral of the Euclidean action $S_E = -i S_L$ instead~\cite{hawking1984}. The amplitude can then be analytically continued back to real time configurations. This approach can also be used for gravity, with a few caveats and subtleties~\cite{hawking1979pathintegralgravity, hawking1983cosmological, hawking1984}. The dominant contributions are then from the metrics which are close to the solutions of the field equations. In particular, when the fields are near their ground state - arguably a good approximation for the present universe - the probability is approximately given by the effective Euclidean action $S_{\text{eff}}$ (i.e. classical, and a solution to the EoM)~\cite{hawking1979pathintegralgravity, hawking1984}. In this case, a probability distribution for $c$ is obtained
\begin{equation}
    P(c) \propto \exp(-S_{eff}(c)),
    \label{p}
\end{equation}
which from (\ref{1b}) produces a sharp probability peak as $\lambda_{eff}(c) \rightarrow 0^+$, thus justifying its vanishing~\cite{hawking1984}.


\subsection{Flaws in the Mechanism and Existing Fixes}

However, this mechanism suffers from two main issues. First, interpreting a path integral over geometries as a probability distribution via a Wick rotation is ill-defined at best~\cite{weinberg1989, Duncan1989, ambjorn1998non, feldbrugge2017lorentzian}. Second, only the EoM for $A$ was solved and substituted into the action, ignoring the EoM for $e^a$. All the EoM should be extracted before any substitutions, since these can alter the resulting EoM. This is true in the present case, as was demonstrated by Duff~\cite{HawkingRetort}, who showed that first extracting the EoM for $A$ \textit{and} $e^a$ before any substitution into the action, one obtains respectively
\begin{equation}
\begin{split}
    & d \ast F = 0, \\
    & \epsilon_{abcd} e^b \left( R^{cd} + \frac{1}{3} e^c e^d \Lambda \right) - \frac{1}{6\kappa} \ast F F_{abcd} e^b e^c e^d = 0,
\end{split}
\label{1e}
\end{equation}
which together yield the same condition as previously ($\ast F = c$), as well as the Einstein field equation with $\Lambda \rightarrow \lambda_{\text{eff}}(c)$
\bea
\label{1c}
& \epsilon_{abcd} e^b \left( R^{cd} + \frac{1}{3} e^c e^d \lambda_{\text{eff}}(c) \right) = 0, \\
\label{1d}
& \lambda_{\text{eff}} (c) = \Lambda - \frac{c^2}{2 \kappa}.
\eea
Therefore, this definition of $\lambda_{\text{eff}} (c)$ is the measurable effective cosmological constant, rather than that defined from the action in ($\ref{1b}$), since it is the EoM that describe the actual Physics. The sign flip in the last expression then invalidates the proposed mechanism, because substituting these solutions back into the action, with the new $\lambda_{\text{eff}}(c)$ definition, the expression becomes
\begin{equation}
    S_{\text{eff}} = - \frac{48 \pi^2 \kappa}{\lambda_{\text{eff}}(c)} + \frac{96 \pi^2 \kappa \Lambda}{\lambda_{eff}(c)^2}.
    \label{2a}
\end{equation}
Clearly the right hand term dominates for small $\lambda_{\text{eff}}(c)$ which, looking also at (\ref{p}), has the wrong sign to exhibit a probability peak.

Nevertheless, this sign problem was later fixed in two possible ways. First, by introducing a suitable boundary term \cite{Duncan1989, wu2008cosmological, aurilia1980hidden}:
\begin{equation}
    S_1 = - 2c \int F \; , \; \; S_2 = - 2 \int d(A \wedge \ast F),
    \label{2}
\end{equation}
with the $c$ here set to be the same as the integration constant above (and so $S_1$ appears unsuitable for this argument, but will be referred to later). On shell these both become $-2 c^2 \int \ast 1$, effectively flipping the $c^2$ term's sign. Second, on a closed manifold use can be made of the Hodge decomposition theorem \cite{Duncan1989} by splitting $F = \tilde{F} + c \ast 1$, where $\tilde{F} = dA$ is an exact form and $c \ast 1$ is obviously a harmonic form. The only term entering the $A$ EoM is $\int \tilde{F} \wedge \ast \tilde{F}$ so that on shell $\ast \tilde{F} = b$, a constant. However due to the uniqueness of the Hodge decomposition, $b = 0$, and so the only term remaining in the effective action and the Einstein equations is the $c^2$ term, this time without a sign flip. Hawking's original result was thus recovered, with the modified $\lambda_{\text{eff}}(c)$ from (\ref{1c}).

The purpose of this paper is to revisit this controversy and see whether other mechanisms can be found resolving the same problem, preferably without sharing the drawbacks of these proposals. We will also investigate whether the Euclidean action prescription can be replaced by something more physical, without destroying the results. 

The plan of this paper is as follows. In section II, a novel action coupling $F$ to $\Lambda$ via a field $\lambda(x)$ is shown to straightforwardly produce a Baum-Hawking-Coleman (BHC) probability peak, while simultaneously being a generalisation of Unimodular Gravity. Section III modifies this action to more closely resemble Hawking's, exhibiting generalised BHC results.
A further generalization and connection is found in Section~\ref{CS}, where the possibility of relating the Chern-Simons density to the Hawking 3-form is shown.
In Section V the dependence of these successes on the dubious probability interpretation based on Euclideanised gravity actions is examined. An alternative probability interpretation, due to Vilenkin, is presented and applied to the theories proposed in this paper, with positive results.  These are then all brought together and discussed in section VI.


\section{$\Lambda - F$ Coupling and Unimodular Gravity}

In this Section we first show how a mechanism similar to Hawking's can be implemented without 
having to appeal to boundary terms or use a Hodge decomposition (which has the drawback of 
introducing in the action parameters fixed to counter others appearing as solutions to the equations
of motion). Our proposal involves a direct coupling of $F$ to $\Lambda$ using the field $\lambda(x)$ as
\begin{eqnarray}
    S &=& \frac{\kappa}{2} \int \epsilon_{abcd} e^a e^b \left( R^{cd} + \frac{1}{6} e^c e^d (\Lambda + \lambda(x)) \right) \nonumber\\
    &&+ \int \sigma(\lambda(x))F.
    \label{3}
\end{eqnarray}
again with a Euclidean action, and $\sigma(\lambda(x))$ so far arbitrary. Varying with respect to $A$, $\lambda(x)$ and $e^a$ leads respectively to
\begin{equation}
\begin{split}
    & d \sigma(\lambda) = 0, \\
    & \epsilon_{abcd} e^a e^b e^c e^d + \frac{12}{\kappa} \sigma'(\lambda) F = 0, \\
    & \epsilon_{abcd} e^b \left( R^{cd} + \frac{1}{3} e^c e^d \Lambda \right) - \frac{1}{6\kappa} \sigma(\lambda) F_{abcd} e^b e^c e^d = 0,
\end{split}
\label{3a}
\end{equation}
where $\sigma'(\lambda) \equiv d\sigma(\lambda)/d\lambda$. The first equation forces $\lambda(x) = c$ to be constant, while the second can be rearranged to $\ast F = - \frac{\kappa}{12 \sigma'(c)}$, so that the Einstein equations are again derived (\ref{1c}) with a different effective cosmological constant
\begin{equation}
    \lambda_{\text{eff}}(c) = \Lambda + c + \frac{\sigma(c)}{\sigma'(c)}.
    \label{3d}
\end{equation}
$\lambda_{\text{eff}}(c)$ can thus be identified from the Einstein equations and substituted back into the effective action under the same closed universe assumption,
leading to:
\begin{equation}
    S_{\text{eff}} = - \frac{48 \pi^2 \kappa}{\lambda_{\text{eff}}(c)} - \frac{96 \pi^2 \kappa}{\lambda_{eff}(c)^2} \frac{\sigma(c)}{\sigma'(c)}.
    \label{4}
\end{equation}
This exhibits the desired probability peak for vanishing $\lambda_{\text{eff}}(c)$ if $\sigma(c) / \sigma'(c) > 0$, without any need for boundary terms.

A notable example for the choice $\sigma(\lambda(x)) \propto \lambda(x)$, is derived in Henneaux and Teitelboim's Hamiltonian analysis of Unimodular Gravity \cite{henneaux1989}. They write the action in terms of the ADM metric, obtaining the EoM by varying all components of this (spacial metric, its conjugate momentum and the lapse function), except for the determinant of the full metric itself. The EoM reproduce General Relativity, with primary, secondary and tertiary constraints. These can all be incorporated directly into the action, and in order to write this as a manifestly local and Lorentz invariant action, the rank 3 tensor field $A_{\mu \nu \lambda}$ is introduced, as the dual to a vector field $\mathcal{T}^{\mu}$ containing a time field $\mathcal{T}^0$ and Lagrange multipliers $\mathcal{T}^i$, as defined in~\cite{henneaux1989}. This can be repackaged into the form
\begin{equation}
    S = \frac{\kappa}{2} \int \epsilon_{abcd} e^a e^b \left( R^{cd} + \frac{1}{6} e^c e^d \tilde{\Lambda} \right) + 2 \kappa \int \tilde{\Lambda} F,\\
    \label{5}
\end{equation}
where $\tilde{\Lambda}$ arises as an integration constant. It is an interesting result that one can arrive at this action from (\ref{3}) by setting $\lambda(x) \rightarrow \lambda$ to a dynamical spacetime constant, and defining 
\bea
\sigma(\lambda) &=& 2 \kappa \tilde{\Lambda}(\lambda)\\
\tilde{\Lambda}(\lambda) &=& \Lambda + \lambda
\eea
Unimodular Gravity could, therefore, provide the motivation for an implementation of a BHC type mechanism, to justify a suppressed cosmological constant.

Recent generalisations to Unimodular Gravity in \cite{barvinsky2017darkness} and particularly \cite{barvinsky2019dynamics} follow a similar process but with a more general constraint, the effects of which are briefly sketched out here for completeness. This leads to a "dark" stress-energy tensor with the form of a perfect fluid, rather than simply a cosmological constant as an integration constant, which is merely a special case here. The trade-off is loss of Lorentz-invariance, though this may well not be an issue, particularly in the context of inflation and the very early universe. One might wonder if the correspondence between our proposed action and Unimodular gravity could be maintained in these generalisations by also doing away with Lorentz invariance - this possibility has not been looked at in detail so far.


\section{$\Lambda - F^2$ Modification}

The parallel with Hawking's original proposal can be more directly seen by modifying the $F$ term in (\ref{3}) to mirror Hawking's:
\begin{equation}
    \int \sigma(\lambda(x))F \rightarrow \int \sigma(\lambda(x)) F \wedge \ast F.
    \label{6}
\end{equation}
The equations of motion are, in the same order as previously
\begin{equation}
\begin{split}
    & d \left(\sigma(\lambda) \ast F\right) = 0, \\
    & \epsilon_{abcd} e^a e^b e^c e^d + \frac{12}{\kappa} \sigma'(\lambda) F \wedge \ast F = 0, \\
    & \epsilon_{abcd} e^b \left( R^{cd} + \frac{1}{3} e^c e^d \Lambda \right) - \frac{1}{6\kappa} \sigma(\lambda) \ast F F_{abcd} e^b e^c e^d = 0,
\end{split}
\label{6a}
\end{equation}
which can be solved in a similar way, though this time the EoM provide an explicit solution for $\sigma$: 
\be\sigma(\lambda(x)) = \frac{1}{a + \frac{4 \kappa \lambda(x)}{b^2}},
\ee
along with the usual Einstein equations (\ref{1c}), and:
\begin{equation}
    \begin{split}
        & \lambda_{\text{eff}}(a, b) = \Lambda - \frac{ab^2}{2 \kappa}, \\
        & \ast F = \frac{b}{\sigma(\lambda(x))},
    \end{split}
\end{equation}
with $a$ and $b$ integration constants. This produces a very similar result to Hawking's original action (\ref{2a}):
\begin{equation}
    S_{eff} = - \frac{48 \pi^2 \kappa}{\lambda_{\text{eff}}(c)} + \frac{96 \pi \kappa \Lambda}{\lambda_{\text{eff}}(c)^2} + \frac{\kappa}{12} \int \ast\lambda(x).
    \label{7}
\end{equation}
Now, one could define $\lambda(x)$ to cancel out the last two terms, but there is no good motivation for this. 

More interestingly, the action can be ``corrected" in the same way as Hawking's, with boundary terms
\begin{equation}
    S_1 = - 2b \int F \; , \; \; S_2 = - 2 \int d \big[ \sigma(\lambda(x)) A \wedge \ast F \big],
    \label{9}
\end{equation}
which reduce to $-2 ab^2 \int \ast 1$ on shell. In this case, $S_1$ might not be discounted so readily, since only $b$ is set explicitly, while $a$ remains unconstrained. Furthermore, the Hodge decomposition argument can also be used for this action, splitting again $F = \tilde{F} + c \ast 1$ with $\tilde{F} = dA$, so that the EoM together require
\begin{equation}
     \ast \tilde{F} = \frac{b}{\sigma(\lambda)} - c.
    \label{8}
\end{equation}
Due to the uniqueness of the Hodge decomposition, this leads to either $c = 0$ and so $\ast F = 0$ (trivial), or $\ast \tilde{F} = 0$. As a result, $\lambda(x)$ drops out of the EoM and effective action entirely (and is in fact also set to a constant), so that equation (\ref{1b}) is recovered, with a different $\lambda_{eff}$:
\begin{equation}
    S_{\text{eff}} = - \frac{48 \pi^2 \kappa}{ \lambda_{\text{eff}}(a, c)} \; , \; \; \lambda_{\text{eff}}(a,c) = \Lambda - \frac{ac^2}{2 \kappa}.
\end{equation}
Either one or two integration constants are thus available to minimize the effective cosmological constant, depending on one's preferred method. $\Lambda - F$ coupling does indeed seem to lead to a clear generalisation of both Unimodular gravity and the Baum-Hawking-Coleman mechanism, through $\Lambda - F^2$ coupling.


\section{Possible relation with Chern-Simons theory}\label{CS}

An interesting further bridge can be found with the alternative replacement:
\begin{equation}
    \int \sigma(\lambda(x))F \rightarrow \int \sigma(\lambda(x)) R \wedge \ast R.
\end{equation}
With this prescription, we are effectively replacing the 3-form appearing in unimodular gravity by the imaginary part of the Chern-Simons density:
\bea
{\cal L}_{CS}&=& {\rm Tr} \left(A_{SD} dA_{SD} +\frac{2}{3} A_{SD} A_{SD}  A_{SD}\right)\nn\\
&=&
-\frac{1}{2} \left( A_{SD}^i dA_{SD}^i +\frac{1}{3}\epsilon_{ijk} A_{SD}^i A_{SD}^j A_{SD}^k\right).\label{CalLCS}
\eea
(where $A_{SD}^i$ is the Self-dual part of the spin-connection).
This is true since:
\be
R \wedge \ast R=4d\Im {\cal L}_{CS},
\ee
(see, e.g.~\cite{CS}).
By choosing:
\be
\sigma=-\frac{3}{2\lambda}
\ee
we are therefore connecting directly with the quasi-topological theory in~\cite{DynL,DynL0}. 
Unsurprisingly, a primary constraint (see~\cite{DynL,DynL0} for details)
then forces the time field ${\cal T}^0$ found in unimodular gravity~\cite{henneaux1989} to be nothing but 
the Chern-Simons time~\cite{Chopin,time}. Recall that the unimodular prescription may be implemented by the addition:
\be
S_0\rightarrow S=S_0 - \int d^4 x \Lambda \partial_\mu T^\mu_\Lambda
\ee
generating a time field:
\be
{\cal T}^0=\int d^3 x T^0.
\ee

We will not explore further this connection in this paper, but stress
that it is possible to keep Hawking's 3-form within the gravitational sector
via this choice. Note that if, on the one hand we have replaced $F$ with $R$ in (\ref{6}), on the other we have ended up with an action of the form
(\ref{3}) with $F=dA$ and the 3-form $A$ given by:
\be
A=\Im {\cal L}_{CS}
\ee
where ${\cal L}_{CS}$ is obtained from the gravitational spin-connection alone.

We note, however, an interesting connection with the problem of time in quantum gravity\cite{Isham,gielen,time}. Here, as in unimodular gravity, the "relational" or "physical" time that converts the WDW equation into a Schroedinger equation is the momentum of the cosmological constant (or a function thereof), rendered a physical variable by the extension of GR contained in the theory. In our case a (first class) constraint forces this momentum to be the Chern-Simons time~\cite{DynL,DynL0} (in unmodular gravity~\cite{henneaux1989}, a Hamilton equation forces this momentum, or rather, its zero mode to be Misner's volume time~\cite{misner}). We note that Chern-Simons time was proposed~\cite{Chopin} as a measure of time long before modern discussions (and indeed it is related to the even earlier York time~\cite{york}).


\section{Probability Flux in FLRW Spacetime}

However, these connections notwithstanding, the issue of using a Euclidean action to describe a gravitational theory on a Lorentzian manifold has yet to be addressed. Here Vilenkin's probability flux may be of use \cite{vilenkin1988quantum}, where he identifies the Wheeler-de-Witt (WdW) equation \cite{dewitt1967quantum, wheeler1968battelle} with the Klein-Gordon (KG) equation, by separating the the Hamiltonian operator into its kinetic and potential terms:
\begin{equation}
    \hat{H} \Psi = \left( a^{-p} \frac{\partial}{\partial a} a^p \frac{\partial}{\partial a} - U(a) \right) \Psi(a) = 0.
    \label{10}
\end{equation}
$a = a(t)$ is the expansion factor from the FLRW metric, and $p$ encodes operator ordering. One can compare this to the KG equation $(\partial_{\mu} \partial^{\mu} - m^2 ) \phi(x) = 0$, which has a conserved current - particle number - given by $j^{\mu} = \frac{1}{2i} \left( \phi^* \partial^{\mu} \phi - \phi \partial^{\mu} \phi^* \right)$. Therefore, Vilenkin similarly defined the conserved current
\begin{equation}
    j(a) = \frac{i}{2} a^p \left( \Psi^*(a) \frac{\partial}{\partial a} \Psi(a) - \Psi(a) \frac{\partial}{\partial a} \Psi^*(a) \right)
    \label{11}
\end{equation}
as the probability flux in superspace. For a potential term of the form $U(a) = A a^2 \left( 1 - B^2 a^2 \right)$ with $A$ and $B$ constants, the WKB solutions for the classically allowed ($a \geqslant B^{-1}$) and classically forbidden ($a < B^{-1}$) regions are given respectively as \cite{vilenkin1988quantum, vilenkin1994approaches}:
\begin{equation}
    \begin{split}
        & \Psi_{\pm}^{(1)}(a) = \text{exp} \left( \pm i \int_{B^{-1}}^a \sqrt{-U(a')} da' \mp \frac{i \pi}{4} \right), \\
        & \Psi_{\pm}^{(2)}(a) = \text{exp} \left( \pm \int^{B^{-1}}_a \big|\sqrt{-U(a')}\big| da' \right),
    \end{split}
    \label{11a}
\end{equation}
assuming a tunnelling boundary condition. If one substitutes these solutions into the probability flux formula (\ref{11}), one obtains
\begin{equation}
    \begin{split}
        & j^{(1)}_{\pm}(a) = \mp a^p \sqrt{-U(a)}, \\
        & j^{(2)}_{\pm}(a) = 0,
    \end{split}
    \label{12}
\end{equation}
so that the former case is dependent upon the potential term $U(a)$.


\subsection{Application to the 3-form}

The objective is then to obtain a semi-classical expression for $U(a)$ from Hawking's action, by performing a Hamiltonian analysis of it in FLRW spacetime. For this section, the Einstein-Hilbert formalism will be used for clarity, and the action will of course be Lorentzian since the whole point is to avoid the Euclidean issues. The total Lagrangian with the field strength tensor $F_{\mu \nu \alpha \beta} = 4 \partial_{[\mu} A_{\nu \alpha \beta]}$ coupled to gravity is
\begin{equation}
    \begin{split}
        & \mathcal{L} = \mathcal{L}_G + \mathcal{L}_F, \\
        & \mathcal{L}_G = \kappa \sqrt{-g} \left( R - 2 \Lambda \right), \\
        & \mathcal{L}_F = - q \sqrt{-g} F_{\mu \nu \alpha \beta} F^{\mu \nu \alpha \beta},
    \end{split}
    \label{13}
\end{equation}
with $R$ the Ricci scalar, and $q > 0$ a constant. Now in an FLRW regime $ds^2 = -N^2 dt^2 + a(t)^2 \gamma_{ij} (\vec{x}) dx^i dx^j$, $\sqrt{-g} \rightarrow N \sqrt{\gamma} a(t)^3$, and the gravitational Lagrangian can written as
\begin{equation}
    \mathcal{L}_G = 6 \kappa \sqrt{\gamma} a \left( - \frac{1}{N} (\dot{a})^2 + Nk - \frac{N}{3} \Lambda a^2 \right),
    \label{14}
\end{equation}
using integration by parts to avoid the $\ddot{a}$ term. Meanwhile, $\mathcal{L}_F$ can be decomposed in the following way:
\begin{equation}
    \begin{split}
        \mathcal{L}_F = & - 4 q N \sqrt{\gamma} a^3 \left( \dot{A}_{ijk} \dot{A}^{ijk} + 3 \partial_i A_{jk0} \partial^i A^{jk0} \right. \\
        & \left. - 6 \dot{A}_{ijk} \partial^i A^{jk0} - 6 \partial_i A_{jk0} \partial^j A^{ik0} \right).
    \end{split}
    \label{15}
\end{equation}
The conjugate pairs that must be considered are $(N, p_N = 0), (\gamma^{ij}, P_{ij} = 0), (A^{0ij}, \pi_{0ij} = 0), (a, p_a), (A^{ijk}, \pi_{ijk})$, so that the first 3 are constraints, $p_a = - \frac{12 \kappa \sqrt{\gamma}}{N} \dot{a}a$ and $\pi_{ijk} = -8q N \sqrt{\gamma} a^3 \left( \dot{A}_{ijk} - 3 \partial_i A_{jk0} \right)$. The Legendre transform is then very similar to the uncoupled case:
\begin{equation}
    \begin{split}
        \mathcal{H} = & \mathcal{H}_G + \mathcal{H}_F, \\
        \mathcal{H}_G = & - \frac{N}{24 \kappa \sqrt{\gamma}} \frac{p_a^2}{a} - 6 \kappa N \sqrt{\gamma} a\left( k - \frac{\Lambda}{3} a^2 \right), \\
        \mathcal{H}_F = & - \frac{1}{16q N \sqrt{\gamma} a^3}\pi_{ijk} \pi^{ijk} + 3 \pi_{ijk} \partial^i A^{jk0} \\
        & - 24 q N \sqrt{\gamma} a^3 \left( \partial_i A_{jk0} \partial^i A^{jk0} - 24 \partial_i A_{jk0} \partial^j A^{ik0} \right),
    \end{split}
    \label{16}
\end{equation}
but with additional constraints to take into account. Now determining the following constraint using integration by parts
\begin{equation}
    \dot{\pi}_{jk0} = 3 \partial^i \pi_{ijk} - 48q \left( \partial^i \partial_i A_{jk0} + \partial^i \partial_j A_{ik0} \right) \approx 0,
    \label{17}
\end{equation}
and substituting $\dot{\pi}_{jk0}$ into the F-sector Lagrangian and Hamiltonian, using integration by parts again, one finds:
\begin{equation}
    \mathcal{H}_F = \mathcal{L}_F.
    \label{18}
\end{equation}
This makes sense since it only has a kinetic term. Note that from the Euler-Lagrange equations for $A_{\mu \nu \alpha}$, $F_{\mu \nu \alpha\ \beta} = c \epsilon_{\mu \nu \alpha \beta}$ implies that semi-classically $\mathcal{L}_F = 24 N \sqrt{\gamma} a^3 q c^2$, and finally:
\begin{equation}
    \mathcal{H}_F = 24 N \sqrt{\gamma} a^3 q c^2.
    \label{19}
\end{equation}
So far this is all the same as the uncoupled case. The first change comes from imposing the $\dot{p}_N \approx 0$ constraint, which when comparing with (\ref{15}) sets $\mathcal{H}_G = \mathcal{L}_F$, and thus from (\ref{18}) $\mathcal{H}_G = \mathcal{H}_F$. Therefore $\mathcal{H} \neq 0$, and for the Hamiltonian constraint to be applied in the WdW equation, it should instead be of the form:
\begin{equation}
    \mathcal{H}' = \mathcal{H}_G - \mathcal{H}_F.
    \label{20}
\end{equation}
Now when integrating over space to obtain a Hamiltonian constraint, there is the subtle issue of the $1 / \sqrt{\gamma}$ factor in the kinetic term of the gravitational Hamiltonian density in (\ref{16}). Strictly speaking, one should integrate the Lagrangian densities, such as (\ref{14}), over space before performing the Legendre transform, to get a Hamiltonian rather than a Hamiltonian density. Therefore the $\sqrt{\gamma}$ term becomes $V_c = \int d \vec{x}^3 \sqrt{\gamma}$, the co-moving volume, in the Lagrangian, and the resulting semi-classical Hamiltonian constraint. Further setting $q = 1 / 4!$ to be consistent with the EC action (\ref{1}), one obtains:
\begin{equation}
    \begin{split}
        & H' = - \frac{N}{24 \kappa V_c} \frac{p_a^2}{a} - 6 \kappa N V_c a \left( k - \frac{a^2}{3} \lambda_{\text{eff}}(c) \right), \\
        & \lambda_{\text{eff}}(c) = \Lambda - \frac{c^2}{2 \kappa}
    \end{split}
    \label{21}
\end{equation}
The effective cosmological constant is exactly the same as that obtained from the working mechanisms earlier, the redefinition of $H \rightarrow H'$ having effectively taken care of the sign flip. Now clearly from (\ref{20}), $\hat{H}' \Psi(a) = 0$ as required, and can be quantized using Dirac's quantization prescription $\{ a, p_a \} = 1 \rightarrow \left[ \hat{a}, \hat{p}_a \right] = i \hbar$, which can be achieved with $\hat{p}_a = - i \hbar \frac{\partial}{\partial a}$. With some choice of $\hat{p}_a$-$\hat{a}^{-1}$ ordering represented by $p = 0, -1$, the resulting WdW equation can be written
\begin{equation}
    \left( a^{-p} \frac{\partial}{\partial a} a^p \frac{\partial}{\partial a} - \frac{144 \kappa^2 V_c^2}{\hbar^2} a^2 \left[ k - \frac{\lambda_{\text{eff}}(c)}{3} a^2 \right] \right) \Psi(a) = 0
    \label{21b}
\end{equation}
The $a \frac{\partial^2}{\partial a^2}(a^{-1})$ case does not seem to be covered by Vilenkin's $p$-ordering, and will be ignored here. However, it appears the arguments below would still likely apply, since it would lead to a similar kinetic term to the $p=-1$ case, with an additional $2a^{-2}$ term added to the potential. Nevertheless, comparing to Vilenkin's expression (\ref{10}), and noting $V_c = V / \left( N \int dt a(t)^3 \right)$, the semi-classical potential term sought after, again for a closed universe solution ($V = 24 \pi^2 / \lambda_{eff}(c)^2$, $k=+1$), is
\begin{equation}
    U(a) = \left(\frac{288 \pi \kappa}{\hbar T \lambda_{\text{eff}}(c)^2} a \right)^2\left( 1 - \frac{a^2}{3} \lambda_{\text{eff}}(c) \right),
    \label{22}
\end{equation}
with $T = \int dt a(t)^3$, so that substituting this into the probability current for the classically allowed region (\ref{12}) yields
\begin{equation}
    j^{(1)}_{\pm}(a,c) = \mp a^{p+1} \frac{288 \pi \kappa}{\hbar T \lambda_{\text{eff}}(c)^2} \sqrt{1 - \frac{a^2}{3} \lambda_{\text{eff}}(c)}.
    \label{23}
\end{equation}
Therefore $|j^{(1)}_{\pm}(c)| \rightarrow \infty$ as $\lambda_{eff}(c) \rightarrow 0^+$, so long as $T$ does not tend to infinity. Note that the condition $\mathcal{H}_G = \mathcal{H}_F$ leads to the first massless Friedmann equation with a modified cosmological constant $\Lambda \rightarrow \lambda_{eff}(c)$, as does the trace of the constraint $\dot{P}_{ij} \approx 0$. The EoM of $p_a$ likewise leads directly to the second massless Friedmann equation with the same modification.


\subsection{Adding matter}

Finally, the behaviour of $T$ can be determined by including some general matter Lagrangian $\mathcal{L}_M(N, a, \gamma_{ij}, \phi) \rightarrow \mathcal{H}_M$, so that the same derivation can be performed, where instead of $\mathcal{H}' = \mathcal{H}_G - \mathcal{H}_F$, one obtains $\mathcal{H}' = \mathcal{H}_G - \mathcal{H}_F + N \sqrt{\gamma}a^3 \rho$. The matter density $\rho$ is naturally determined from the resulting Friedmann equations, along with the pressure-density ratio $w$:
\begin{equation}
    \begin{split}
        & \left( \frac{\dot{a}}{a} \right)^2 + N^2 \frac{k}{a^2} = \frac{N^2}{3} \left( \frac{\rho}{2 \kappa} + \lambda_{\text{eff}}(c) \right) , \\
        & \frac{\ddot{a}}{a} = - \frac{N^2}{3} \left[ \frac{\rho}{4 \kappa} \left( 1 + 3w \right) - \lambda_{\text{eff}}(c) \right] , \\
        & \rho = \frac{1}{\sqrt{\gamma} a^3} \frac{\partial \mathcal{H}_M}{\partial N} \; , \; \;
        w = - \frac{1}{3} \left( 1 + \frac{a^3}{N} \frac{\partial \mathcal{H}_M / \partial a}{\partial \mathcal{H}_M / \partial N} \right).
    \end{split}
    \label{23b}
\end{equation}
These equations can then be solved by moving to conformal time $N dt \rightarrow a(\eta) d \eta$, $a(t) \rightarrow a(\eta)$, and absorbing $\lambda_{\text{eff}}(c)$ into $\tilde{\rho} = \rho + 2\kappa \lambda_{\text{eff}}(c)$ and $\tilde{w} = w / \left( 1 + 2 \kappa \lambda_{\text{eff}}(c) / \rho \right)$, which then simply replace $\rho$ and $w$ above. Now, defining $C = \left( 1 + 3 \tilde{w} \right) / 2$, multiplying this to the first equation and then adding the result to the second, while defining a new function $a' / a = f' / C f$ (prime denotes derivative with respect to $\eta$), reduces this to:
\begin{equation}
    \frac{f''}{f} + C k = 0.
    \label{23c}
\end{equation}
Assuming $C$ remains constant, enforcing the boundary condition $f(0) = 0$, and converting back to $a$, one finds $a(\eta) = a_{max} sin(\sqrt{C} \eta) ^{\frac{1}{C}}$. $a_{max}$ can then be determined by substituting the solution back into the first Friedmann equation in conformal time, and choosing $\eta = \pi / 2 \sqrt{C}$, so that
\begin{equation}
    \begin{split}
        & a(\eta) = \pm \sqrt{\frac{3}{\lambda_{\text{eff}}(c) + \rho / 2 \kappa}} sin(\sqrt{C} \eta) ^{\frac{1}{C}}, \\
        & C = \frac{1}{2} \left( 1 + \frac{3 w}{1 + 2 \kappa \lambda_{\text{eff}}(c) / \rho} \right),
    \end{split}
    \label{23d}
\end{equation}
where as usual $w = 0, 1/3$ for a cold matter or radiation dominated universe respectively. Note that the $w = -1$ case is absent since any $\lambda_{eff}(c)$ dependence has been extracted from $w$. In conformal time, $T$ becomes $T = \frac{1}{N} \int d\eta a(\eta)^4$, and changing variables again to $x = \sqrt{C} \eta$, and setting the limits to when $a(\eta) = 0$, this is
\begin{equation}
    T = \frac{9}{N \sqrt{C} \left(\lambda_{\text{eff}}(c) + \rho / 2 \kappa \right)^2}  \int_0^{\pi} dx \text{sin}(x)^{\frac{4}{C}}.
    \label{23e}
\end{equation}
For $w = 0$, $4/C = 8$, which can be readily solved for:
\begin{equation}
    T_0 = \frac{9 \sqrt{2}}{N \left( \lambda_{\text{eff}}(c) + \rho / 2 \kappa \right)^2} \frac{35 \pi}{128}.
    \label{24}
\end{equation}
However, the $w = 1/3$ cannot be solved as $C$ still depends on the specifics of $\lambda_{\text{eff}}(c) / \rho$. To get around this, note the behaviour of (\ref{23e}), as $\lambda_{\text{eff}}(c) \rightarrow 0^+$, $C \rightarrow \left( 1 + 3 w \right) / 2$, so that the integrals for $w = 0, 1/3$ can both be performed:
\begin{equation}
    \begin{split}
        & T_0 \rightarrow \frac{9 \sqrt{2}}{N \left(\rho / 2 \kappa \right)^2} \frac{35 \pi}{128} \\
        & T_{1/3} \rightarrow \frac{9}{N \left(\rho / 2 \kappa \right)^2} \frac{3 \pi}{8}.
    \end{split}
    \label{25}
\end{equation}
This result for $T_0$ agrees with that obtained previously (\ref{24}) for $\lambda_{eff}(c) \rightarrow 0^+$. In both cases, $T$ does not blow up to infinity, and taking into account the new $\rho$ term in the potential:
\begin{equation}
    j^{(1)}_{\pm}(a,c) = \mp a^{p+1} \frac{288 \pi \kappa}{\hbar T \lambda_{\text{eff}}(c)^2} \sqrt{ 1 - \frac{a^2}{3} \left(\lambda_{\text{eff}}(c) + \frac{\rho}{2\kappa} \right)},
    \label{26}
\end{equation}
it does indeed appear that $|j^{(1)}_{\pm}(c)| \rightarrow \infty$ as $\lambda_{eff}(c) \rightarrow 0^+$.


\section{Discussion and outlook}

It is clear that the BHC mechanism can be made to work with a variety of approaches, as has been shown in previous work~\cite{Duncan1989, wu2008cosmological, aurilia1980hidden}. This paper provides potentially new approaches via new actions, such as $\Lambda - F$ coupling, which exhibits BHC behaviour while simultaneously encompassing a form of Unimodular Gravity presented by Henneaux and Teitelboim~\cite{henneaux1989}. Meanwhile, a variation on this, $\Lambda - F^2$ coupling, demonstrates a clearer generalisation of Hawking's original action, and of the BCH mechanism in general. Therefore, Unimodular Gravity could provide a potential motivation for a BCH-type result. At the very least, the link is noteworthy.

Lastly, using Vilenkin's probability flux~\cite{vilenkin1988quantum} instead of the troublesome Euclidean action, broad agreement was found with the latter's results. This is very encouraging, but one must take note of the assumptions and approximations made. First, the Vilenkin wavefunctions (\ref{11a}) used to derive an expression for the probability flux are WKB approximate solutions to the WdW equation. Fortunately, taking $\lambda_{\text{eff}}(c) \rightarrow 0^+$ makes $U(a)$ tend to infinity, likely leading to a slowly evolving solution, which the WKB approximation was designed for ($U >> K$, for $K$ the kinetic term). Secondly, only the classically allowed region is being considered here, where $a < \sqrt{3 / \left(\lambda_{\text{eff}}(c) + \rho / 2 \kappa \right)}$. Thirdly, considering a semi-classical Hamiltonian naturally leads to its own concerns over potential inaccuracies, which may be partly alleviated by only considering the classically allowed region. Lastly, it was assumed that $C$ in (\ref{23d}) was constant to derive the behaviour of $a(\eta)$, and by extension the matter density $\rho$ was also assumed to be constant, or at least slowly evolving. While this might be compatible with a slowly evolving WKB solution, it may be increasingly inaccurate when integrating over the universe's lifetime, as was done here. Nevertheless, the results of this integration, and indeed the whole probability flux are merely instructive, to gauge the general dependence of this on the effective cosmological constant as it tends to zero.

This probability flux approach can also be applied to the new ``coupling" actions with minor modifications, and is expected to produce broadly similar results. How one would approach quantizing these new variations in the scheme of larger quantum gravity theories remains to be seen - as was done for the original BHC mechanism in the context of String Theory \cite{bousso2000quantization}. All together, the linking of Unimodular Gravity to the BHC mechanism via new actions, with a promising new probability argument, that all appear in agreement, presents at the least an interesting new take on the topic.

We close by noting that connections may probably be found with other work.
Hawking's approach has clear parallels with that of
sequestration~\cite{Sequester} and wormholes~\cite{Coleman}. In all of these approaches
one generates an ensemble of effective field theories with different 
cosmological constants (although they differ in how to select among them). 
Could the lessons learnt in this paper illuminate the differences and interconnections between all of these different mechanisms?
On a different front,
Lambda may be released from constancy in quasi-topological theories~\cite{DynL0,DynL}. Then, Lambda appears closely related to a form of conformal invariance, associated with the presence of torsion~\cite{MZ}, which can be broken together with parity. The quantum cosmology of such theories~\cite{MSSdynL} leads to an approach almost orthogonal to that followed in this paper, which does not mean that they cannot be combined. The possibility that Lambda is a purely quantum phenomenon~\cite{QuantL} in theories where gravity emerges as a Fermi liquid~\cite{Fermi1,Fermi2,Fermi3} is another puzzling connection that remains to be made. Lastly, note that \cite{carroll2017nonlocal} similarly drew links between the BHC mechanism and Unimodular gravity, in the context of imposing a nonlocal constraint on the gravity action.

{\it Acknowledgments}: We would like to thank Antonio Padilla for extensive interaction and suggestions at an early stage of this project, as well as Stephon Alexander for bringing Hawking's original paper to our attention. This work was supported by the STFC Consolidated Grant ST/L00044X/1 (JM).  

\bibliography{MScReport} 

\end{document}